\documentclass[a4paper,preprint]{revtex4}
\usepackage{hyperref}
\usepackage{graphicx}
\usepackage{url}
\usepackage{longtable}
\usepackage{amsmath}

\def \gsim{\mathrel{\mathpalette\@versim>}}
\def \lsim{\mathrel{\mathpalette\@versim<}}
\def\etm{E{\!\!\!/}_T}
\begin{document}
\title{CP-violating HWW couplings at the Large Hadron Collider}
\author{Nishita Desai} 
\affiliation{Harish-Chandra Research Institute,
  Jhunsi, Allahabad - 211 019, India} 
\author{Dilip Kumar Ghosh}
\affiliation{Department of Theoretical Physics, \\ 
Indian Association for Cultivation of Science, 
2A \& 2B Raja S.C. Mullick Road, Kolkata - 700032 , India} 
\author{Biswarup Mukhopadhyaya}
\affiliation{Harish-Chandra Research Institute, Jhunsi, Allahabad -
  211 019, India}
\preprint{HRI-RECAPP-2011-002}
\begin{abstract}
We investigate the possibility of probing an anomalous CP-violating
coupling in the HWW vertex at the LHC.  We consider the production of
the Higgs in association of a W and then decay via the $H \rightarrow
WW$ channel taking into account the limits on the Higgs production
cross section from the Tevatron.  We select the same-sign dilepton
final state arising from leptonic decays of two of the three Ws and
apply cuts required to suppress the standard model background. Several
kinematical distributions and asymmetries that can be used to
ascertain the presence of a non-zero anomalous coupling are presented.
We find that, for Higgs mass in the range 130-150 GeV and anomalous
couplings allowed by the Tevatron data, these distributions can be
studied with an integrated luminosity of 30-50 fb$^{-1}$ at the 14 TeV
run.  Attention is specifically drawn to some asymmetries that enable
one to probe the real and imaginary parts (as well as their signs) of
the anomalous coupling, in a complementary manner.  We also explicitly
demonstrate that showering and hadronisation do not affect the utility
of these variables, thus affirming the validity of parton level
calculations.
\end{abstract}

\maketitle

\section{Introduction}
The most well-motivated explanation for electroweak symmetry breaking
is via the Higgs mechanism.  Although the Higgs boson remains the only
unobserved particle in the Standard Model (SM), there are both
experimental and theoretical bounds on its mass.  The LEP bound of
$114$~GeV has now been supplemented with the Tevatron bounds which
rule out Higgs masses between $158-173$~GeV
\cite{:2010ar,Baglio:2011wn}.

Here we consider the possibility of the Higgs boson existing somewhere
between $130-150$~GeV where the decay width of $H \rightarrow WW$ is
appreciable.  This is the range which has not yet been ruled out by
the Tevatron and is likely to be probed at the earliest at the Large
Hadron Collider (LHC)\cite{Ball:2007zza,Aad:2009wy}.  On the one hand,
there is substantial rate of production; on the other, the viability
of the $WW^*$ decay channel avoids the requirement of the two-photon
mode, and consequently the requirement of a large integrated
luminosity for discovery.

In such a situation, we wish to probe whether the $HWW$-coupling is
purely described by the standard model.  Such couplings can be probed
in the relatively clean leptonic channels and previous studies for
$HWW$ and $HZZ$ couplings at the LHC can be found in
\cite{Zhang:2003it, Biswal:2005fh, Godbole:2007cn, Qi:2008ex,
  Zhang:2008zzw, Han:2009ra, Christensen:2010pf}.  Many studies for
both $HWW$ and $HZZ$ anomalous couplings also exist in the context of
a future $e^+e^-$ collider \cite{Miller:2001bi, Choi:2002jk,
  Rindani:2009pb,Biswal:2008tg, Biswal:2009ar, Takubo:2010tc},
$e\gamma$ collider \cite{Choudhury:2006xe} and photon collider
\cite{Han:2005pu,Sahin:2008qp}.  LEP limits on anomalous Higgs
couplings can be found in \cite{Acciarri:2000ep}.

The primary production channel at the LHC is through gluon-gluon
fusion and would in principle be the cleanest to probe the $HWW$
vertex.  However, the decay $H \rightarrow W^+ W^-$ leads to an
opposite-sign dilepton signature which is prone to large backgrounds
from $pp \rightarrow W^+W^-$.  This background is generally eliminated
by removing back-to-back leptons with an appropriate
cut\cite{Dittmar:1996ss}.  However, these cuts are no longer useful
when one wishes to probe the presence of anomalous couplings because
the difference made by the anomalous couplings in angular
distributions of dilepton events is often in this kinematic region.
We therefore choose to probe the associated production channel
instead.  A study for probing the anomalous couplings in the
vector-boson fusion channel can be found in \cite{Hankele:2006ma}.
One could also consider Higgs production via $pp \rightarrow ZH$.
However, if the $HWW$ vertex has anomalous couplings, it would be
natural to expect the $HZZ$ vertex to also have such couplings.  In
that case, one is left to disentangle the interference of both $HWW$
and $HZZ$ vertices and this will further complicate the study of the
$HWW$ interaction.

Thus we explore the production of the Higgs via $pp \rightarrow WH$,
and its subsequent decay, again through the $HWW$ coupling. The
interplay of anomalous coupling in both the production and decay
vertices makes the resulting phenomenology richer and more
complicated, but free from contamination from other effects.  The
environment of a hadron collider and the presence of two neutrinos in
the final decay products makes the reconstruction of the event and the
extraction of a non-standard $HWW$ vertex difficult.  However, as we
shall see, there are significant differences in angular distributions
which may point to the presence of anomalous contributions.  We will
also specifically address the issue of effect of initial and final
state radiation on the variables as this is a fundamental concern at
the LHC.

The paper is organised as follows. In the next section, we acquaint
the reader with the anomalous couplings, and go on to discuss
model-independent strategies for probing the CP-violating anomalous
coupling, in a parton level Monte Carlo approach. Our event selection
criteria are also discussed there. Section \ref{sec:numerics} contains
our numerical results, including various distributions and asymmetries
relevant for the analysis. In section \ref{sec:ISR}, we report the
results of a study where hadronisation and initial and final state
radiation are included, and try to convince the reader that these do
not alter the conclusions of a parton level study in most cases. Our
conclusions are presented in section \ref{sec:conclusions}.
\section{The anomalous coupling and its simulation}
\label{sec:theo}

The $HWW$ vertex can receive corrections from higher dimensional
operators like $\frac{(\Phi^\dag \Phi)}{\Lambda^2} W_{\mu \nu} W^{\mu
  \nu}$ and $\frac{(\Phi^\dag \Phi)}{\Lambda^2} W_{\mu \nu} \tilde
W^{\mu \nu}$.  The general $HWW$ vertex may then be written in a
model-independent way as $\Gamma_{\mu \nu} W^\mu W^\nu H$ where:
\begin{widetext}
\begin{equation}
\label{eqn:vertex}
\Gamma_{\mu \nu} = \frac{ig M_W}{2}\left(a g_{\mu \nu} 
                + \frac{b}{M_W^2} (p_{1\mu} p_{2\nu} + p_{1\nu}p_{2\mu} - (p1 \cdot p2) g_{\mu \nu}) 
                + \frac{\tilde b}{M_W^2} \epsilon_{\mu \nu \rho \sigma} p_1^{\rho} p_2^{\sigma} \right)
\end{equation}
\end{widetext}
where $p_1$ and $p_2$ are the momenta of the two gauge bosons.  For
this study, we assume a completely phenomenological origin of $b$ and
$\tilde b$.  The Standard Model vertex then corresponds to $b=0$,
$\tilde b = 0$ and $a=1$.  We particularly wish to investigate the
effect of non-zero values of $\tilde b$ which would lead to
$CP$-violation.  Therefore, we set $b$ to zero all along.  We also
include the possibility of a complex $\tilde b$, arising out of some
absorptive part in the effective interaction.

\subsection{Simulation}

To start, a parton-level Monte Carlo analysis has been performed for
investigating the kinematical consequences of the CP-violating
anomalous vertex at the LHC, using leptons in the final state.  In
section \ref{sec:ISR}, we will show that the results of this
simplified analysis are not altered by showering and hadronisation
effects.  We factorize the entire matrix element into two pieces $pp
\rightarrow H \ell \nu (\ell = e, \mu)$and $ H \rightarrow WW^*
\rightarrow \ell \nu \bar f f'$\cite{Arens:1994nc}.  Since the Higgs
is a scalar, we expect that this does not affect the spin
correlations.  Both matrix elements have been calculated using Form
\cite{Vermaseren:2000nd}.  For the first part of our study, we perform
a simple smearing of the lepton momenta to approximate detector
effects with a Gaussian of width given by $\sigma(E) = aE + b\sqrt E$
with $a=0.02$ and $b=0.05$\footnote{B.\ Mellado; private
  communication}.  The lepton identification efficiency has been
assumed to be $100\%$.

We present our calculations for a proton-proton centre-of-mass energy
of 14~TeV.  The signal rates are too small at 7~TeV to be accessible
at the current run with the projected luminosity. The cross section is
calculated using the CTEQ6L1 parton distribution
functions\cite{Pumplin:2002vw} using with the renormalisation and
factorisation scales both set at $\sqrt{\hat s}$, the subprocess
centre-of-mass energy.

The modification of the leading order (LO) decay width in the $ H
\rightarrow f \bar f' \bar f f'$ channel has also been calculated and
taken into account in each case.  We focus on same-sign dileptons
(SSD), when only one of the $W$s from the Higgs decays leptonically.
It is less profitable to look into exclusive opposite-sign dilepton
because of the large background from $W^+W^-$ production.  In Higgs
searches, this background is generally suppressed using a cut on the
angle between the two leptons --- the ones from $W^+W^-$ production
are mostly back-to-back whereas those from Higgs decay are highly
collimated.  Since the excess events due to the term proportional to
$\tilde{b}$ tend to also increase the angle between the opposite-sign
leptons, we cannot really use this as a criterion for cutting out the
background.  We can also look at trilepton states when both the $W$s
decay into leptonic final states but we omit them for this work due to
very low cross sections.

The Tevatron has certain bounds on the cross section of Higgs
production.  The latest CDF bounds in the SSD channel with 7.1
fb$^{-1}$ data on the ratio of the Higgs production cross section to
the SM rate in the electron and muon channels are 9.63 (4.99) on Higgs
masses of 130 (150) GeV\cite{CDF:conf2011}.  The combined CDF and
{D\O} results\cite{:2010ar} put a much stronger upper bound on the
Higgs cross sections by combining various channels.  However, the
anomalous coupling affects only the production via the associated $WH$
production for which the bounds are not as strong.  We present the
results in our paper for a value $|\tilde b| = 0.2$ which satisfies
the above CDF bounds.

\subsection{Backgrounds and Cuts}

At the LHC, the largest contribution to the background for SSD comes
from semileptonic $B$-meson decays in $b \bar b$ production where one
of the $B$-mesons oscillates into its charge conjugate state.  It has
been well-known for some time that the isolation cuts alone are not
enough to suppress this background\cite{Sullivan:2010jk} but an
additional cut on the transverse momentum ($p_T$) is required.  We
found that demanding an additional $p_T$-cuts along with a cut on
missing transverse energy ($\etm$) is very effective for suppressing
this background.  We require two isolation cuts on the leptons,
viz. the sum of $p_T$ of all particles within a cone of 0.2 around the
lepton should be less than 10 GeV and the separation from the nearest
jet should be less than 0.4.  However, these cuts are only fully
relevant after parton showering and hadronisation and therefore will
be considered in detail in section \ref{sec:ISR}.  Therefore, the set
of cuts used for the parton-level analysis are:
\begin{enumerate}
   \item  Lepton rapidity : $\mid \eta \mid <$ 2.5 
   \item  Minimum transverse momenta of the hardest and
                  second hardest leptons : $pT(\ell_1) >$ 40 GeV and 
                  $p_T(\ell_2) >$ 30 GeV respectively
   \item  Missing transverse energy: $\etm >$ 30 GeV
\end{enumerate}

These cuts suppress the $b\bar b$ background completely and reduce the
contribution of $Zb\bar b$, $Wb\bar b$ to very small amounts.  The
$t\bar t$ background is still in the range of several femtobarns and
can be further suppressed using a veto on b-tagged jets and also
restricting the number of hard jets in the final state.  Since both
these cuts are dependent on showering and hadronisation effects, we
shall examine them only in section \ref{sec:ISR}.

\begin{figure}[htb]
\includegraphics[width=80mm]{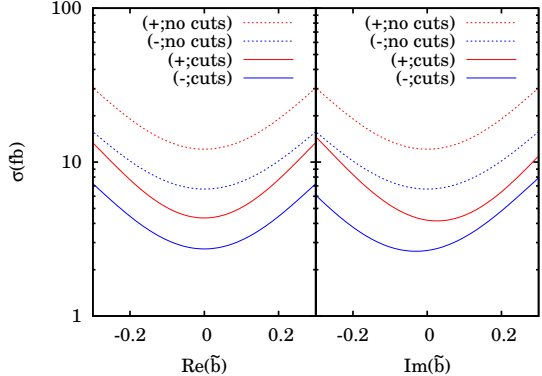}

\caption{\label{fig:crossx-ssd} The behaviour of the cross section
  ($m_H=150$~GeV) in the SSD channel for different values of ${\rm
    Re}(\tilde b)$ and ${\rm Im}(\tilde b)$ (right) and the final
  cross section after all the cuts.  ${\rm Im}(\tilde b)$ is set to
  zero in the left panel and ${\rm Re}(\tilde b)$ is zero in the
  right.  The dashed line with label ``nocuts'' refers to the cross
  section before any cuts are applied whereas the solid lines
  correspond to the cross section after cuts.  The signs $\pm$ refer
  to the charge of the SSD.}
\end{figure}

The effect of cuts for SSD on different values of $\tilde b$ can be
seen in Figure \ref{fig:crossx-ssd}.  We change only one out of ${\rm
  Re}(\tilde b)$ and ${\rm Im}(\tilde b)$ at a time.  Changing ${\rm
  Re}(\tilde b)$ increase the $p_T$ of the leptons however, this
increase is similar for both ${\rm Re}(\tilde b)>0$ and ${\rm
  Re}(\tilde b)<0$.  The case for non-zero ${\rm Im}(\tilde b)$
however, is different.  ${\rm Im}(\tilde b)<0$ enhances the $p_T$ for
the lepton from $W^+$ whereas ${\rm Im}(\tilde b)>0$ enhances the
$p_T$ for the lepton from $W^-$.  This causes an asymmetry in the
cross section after the cuts even though there is no asymmetry to
start with.  This is illustrated in Table~\ref{tab:cutflow-ssd} where
we present the cut flow table for the SM case and for ${\rm Im}(\tilde
b) = \pm 0.2$ for both $\ell^+\ell^+$ and $\ell^-\ell^-$ final states.
The corresponding cross sections with ${\rm Re}(\tilde b)=0.2$ for
$m_H=150(130)$ GeV are (7.64)3.49 fb for $\ell^+\ell^+$ and 4.35
(2.08) fb for $\ell^-\ell^-$.
\begin{table}
\begin{tabular}{|c|ccc|ccc|}
\hline
$m_H$       &          & $130$~GeV  &        &         &  $150$~GeV  &    \\
\hline
$(\tilde b;\pm)$  
             & Cut 1  & Cut 2  &  Cut 3      &  Cut 1&  Cut 2  & Cut 3   \\
\hline
$(0.0;+) $   & 3.80    & 1.56    & 1.49     & 5.97   &  3.06   & 2.99    \\
$(0.0;-) $   & 3.09    & 1.11    & 1.06     & 4.53   &  2.08   & 2.02    \\
$(0.2i;+)$  & 7.69    & 2.81    & 2.77       & 13.86  &  6.21   & 6.16    \\
$(0.2i;-)$  & 5.03    & 2.44    & 2.30       & 8.38   &  4.97   & 4.77    \\
$(-0.2i;+)$ & 7.15    & 2.81    & 2.77       & 12.87  &  8.66   & 8.29    \\
$(-0.2i;-)$ & 5.28    & 1.74    & 1.71       & 8.75   &  3.63   & 3.59    \\
\hline
\end{tabular}
\caption{\label{tab:cutflow-ssd} The effect of cuts on the SSD cross
  section for non-zero ${\rm Im}(\tilde b)$; the $\pm$ signs refer to the
  charge of the SSD.  The cross sections are in $fb$ and are evaluated
  at $\sqrt{s}=14$ TeV The cuts are explained in the text.}
\end{table}

\section{Numerical Results}
\label{sec:numerics}

After applying the cuts described in the previous sections, we are
left with a fairly pure sample of events.  Therefore we shall present
the distributions for signal events only.  Since the strength of the
cross section for different values of the anomalous coupling are
already given in Figure \ref{fig:crossx-ssd}, we will be presenting
only the normalised distributions for the rest of this work.  We also
present distributions only for Higgs mass ($m_H$) of 150~GeV since the
cross section in this case is larger.  The distributions for
$m_H=$ 130~GeV are qualitatively similar.  The asymmetry distributions
are shown for both Higgs masses and it will be seen that $m_H=$130~GeV
is in fact more sensitive to some of them.

\begin{figure}[htb]
\includegraphics[width=80mm]{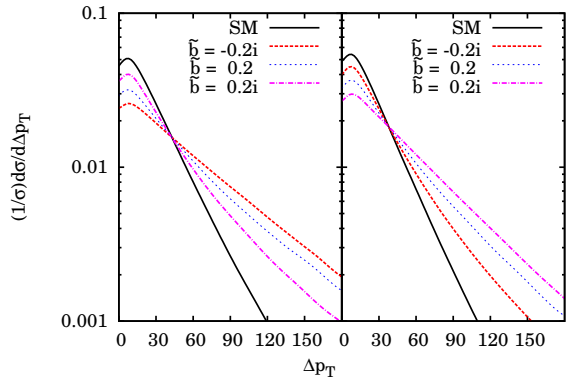}
\caption{\label{fig:pt-ssd} The normalised $\Delta p_T$ distribution
  between two same-sign leptons for $m_H=150$~GeV.  The left(right)
  panel corresponds to $\ell^+\ell^+$ ($\ell^-\ell^-$)-type process.}
\end{figure}

The first variable of interest is the difference in transverse momenta
of the leptons.  The two leptons in the SSD channel are labeled in
descending order of their $p_T$.  We then define $\Delta p_T =
p_T^{(1)} - p_T^{(2)}$.  The charge of the SSD points out whether we
have a $W^+$ or $W^-$ initiated process.  Figure~\ref{fig:pt-ssd}
shows the distribution for both $W^\pm$-type processes.  The sign of
${\rm Re}(\tilde b)$ does not affect the hardness of the distribution.
Therefore, we show only one curve corresponding to ${\rm Re}(\tilde
b)=0.2$.  However, the difference due to change in sign of ${\rm
  Im}(\tilde b)$ is reflected in the two curves corresponding to ${\rm
  Im}(\tilde b)=\pm 0.2$.

\begin{figure}[htb]
\includegraphics[width=80mm]{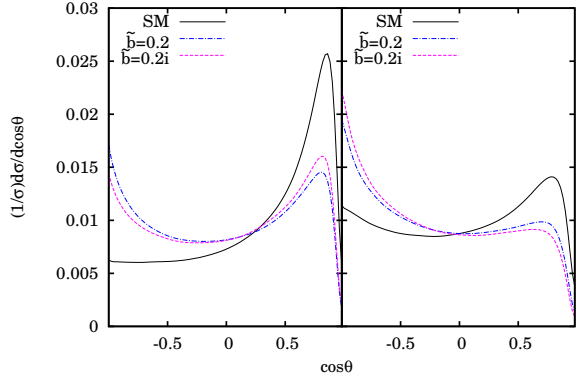}
\caption{\label{fig:costh-ssd} The normalised distribution of cosine
  of the angle between two same-sign leptons for $m_H=150$~GeV.  The
  effect of changing either ${\rm Re}({\tilde b})$ or ${\rm Im}(\tilde b)$ is to
  enhance the back-to-back nature of the leptons.  The left panel
  shows the $\ell^+\ell^+$ whereas the right panel shows the $\ell^-\ell^-$-type
  distributions.}
\end{figure}

Next, we consider the distribution in the angle between the two
same-sign leptons.  In the absence of any cuts, the distribution peaks
at $\theta=0$, i.e. $\cos \theta=1$.  However, the $p_T$ cuts remove
nearly all these highly collinear events.  The peak for SM curve is
shifted from $\cos \theta=1$ to $\cos \theta\sim 0.5$.  Figure
\ref{fig:costh-ssd} shows how the distribution changes for non zero
${\rm Re}(\tilde b)$ and ${\rm Im}(\tilde b)$.  The effect of the
anomalous coupling is to enhance the back-to-back nature of the
distribution.  The forward peak is almost completely diminished.  A
quantitative measure of this change can be made by measuring the
asymmetry around $\cos \theta =0$.

\begin{figure}[htb]
\includegraphics[width=80mm]{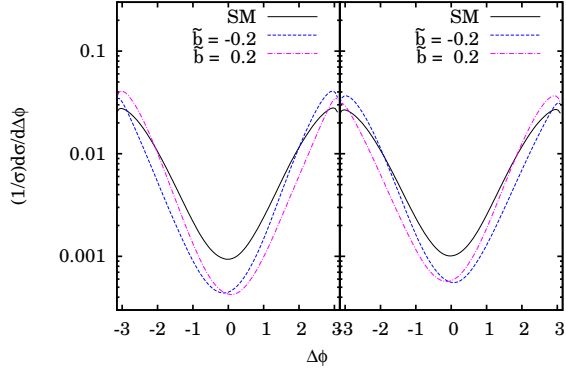}
\caption{\label{fig:dphi-ssd} Effect of positive (dashed) and negative
  (dot-dashed) values of ${\rm Re}(\tilde b)$ on the $\Delta \phi$
  distributions for $m_H=150$~GeV.  The left panel corresponds to
  $\ell^+\ell^+$ and the right to $\ell^-\ell^-$ final states.}
\end{figure}

We then look at the $\Delta \phi$ distribution, where $\Delta \phi$ is
defined as $\phi_{\ell_1} - \phi_{\ell_2}$ and $\phi$ stands for the
azimuthal angle.  In this case however, we adopt a different ordering
of the leptons.  We wish to identify which lepton is more likely to
come from Higgs decay ($\ell_2$) and which from the main hard
interaction ($\ell_1$).  Since one of the $W$s from the Higgs decays
into jets, we would expect the lepton from Higgs decay to be closer to
at least one of the jets than the other lepton.  We therefore pick the
lepton with the smallest distance to any of the jets as $\ell_2$ and
then construct $\Delta \phi$.  Contrary to the previous distributions,
this distribution is particularly sensitive to the sign of ${\rm
  Re}(\tilde b)$ but not to the sign of ${\rm Im}(\tilde b)$.  The
effect of different ${\rm Re}(\tilde b)$ on both $\ell^+\ell^+$ and
$\ell^-\ell^-$ can be seen from Figure~\ref{fig:dphi-ssd}.  A non-zero
${\rm Im}(\tilde b)$ only changes the height of the dip and the
distribution is symmetric about $\Delta \phi = 0$ whereas flipping the
sign of ${\rm Re}(\tilde b)$ flips the distribution as well.

Since the $\Delta \phi$ distribution has a central dip and also shows
left-right symmetry for the standard model case, we can construct two
kinds of asymmetries, viz.
\begin{eqnarray}
A_{SSD1}& = &\frac{\sigma(\Delta \phi>0) - \sigma(\Delta
  \phi<0)}{\sigma(\Delta \phi>0) + \sigma(\Delta \phi<0)}\\ 
A_{SSD2}& = &\frac{\sigma(|\Delta \phi|<\pi/2) - \sigma(|\Delta
  \phi|>\pi/2)}{\sigma(|\Delta \phi|<\pi/2) + \sigma(|\Delta
  \phi|>\pi/2)}
\end{eqnarray}

\begin{figure}[htb]
\includegraphics[width=80mm]{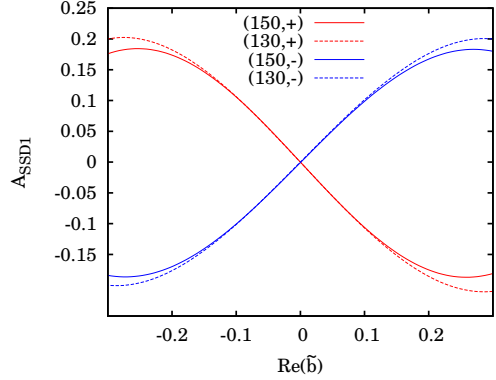}
\caption{\label{fig:asym-dphi-lr}The asymmetry $A_{SSD1}$ for
  different values of ${\rm Re}(\tilde b)$ for $\ell^+\ell^+$(red) and
  $\ell^-\ell^-$(blue) with ${\rm Im}(\tilde b)=0$.  The labels refer
  to the Higgs mass and the sign of the SSD.}
\end{figure}

\begin{figure}[htb]
\includegraphics[width=80mm]{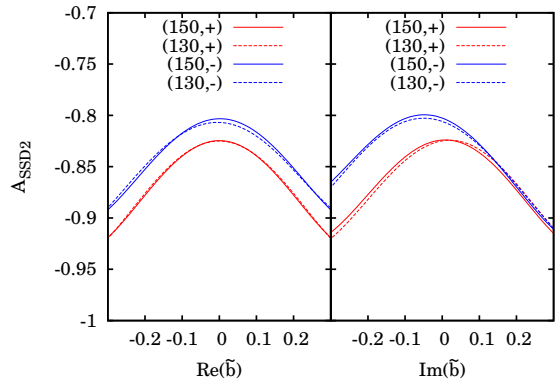}
\caption{\label{fig:asym-dphi-c} The asymmetry $A_{SSD2}$ for
  different values of the anomalous couplings.  The labels indicate
  the Higgs mass and the sign of the SSD.}
\end{figure}

The first is a left-right asymmetry which captures the change in the
sign of ${\rm Re}(\tilde b)$ but remains unaffected by ${\rm
  Im}(\tilde b)$.  The effect of ${\rm Re}(\tilde b)$ on $A_{SSD1}$ is
shown in Figure~\ref{fig:asym-dphi-lr}.  The sign of the asymmetry is
oppositely correlated to the sign of the coupling.  We also look at
$A_{SSD2}$ distribution given in Figure~\ref{fig:asym-dphi-c} which
describes how central the $\Delta \phi$ distribution is.  We notice
that the effect of both ${\rm Re}(\tilde b)$ and ${\rm Im}(\tilde b)$
is similar in this regard.  Therefore if $A_{SSD2}$ shows a
significant deviation from the SM value but $A_{SSD1}$ does not, it
would point to the presence of a non zero ${\rm Im}(\tilde b)$.

For a reasonable estimation at the LHC, we require that the
asymmetries be reasonable separated from the SM value by at least
three standard deviations.  Using the formula in equation (5), for a
value of ${\rm Re}(\tilde b)=0.2$ and $m_H=150$ GeV for
$\ell^+\ell^+$-type events, we find that a luminosity of 30fb$^{-1}$
gives an asymmetry $A_{SSD1}=-0.210\pm 0.065$ and $A_{SSD2}=-0.886\pm
0.031$, both of which are inconsistent with the SM values of
$A_{SSD1}=-0.002$ and $A_{SSD2}=-0.786$ by the required factor.  A
5$\sigma$ difference can be achieved with 50 fb$^{-1}$ data.  The
corresponding $3\sigma$ measurement for $m_H=130$ GeV can be done with
50 fb$^{-1}$ giving $A_{SSD1}=-0.222\pm 0.074$ and $A_{SSD2}=-0.88\pm
0.036$.  A $5\sigma$ measurement would require 140 fb$^{-1}$.

To complement the $\Delta \phi$ variable which is sensitive to the
sign of ${\rm Re}(\tilde b)$, we would also like to construct a
variable that is sensitive to the sign of ${\rm Im}(\tilde b)$.  We
first reconstruct the $W$ that has decayed into jets and obtain its
rapidity, $\eta_W$.  We then construct $\Delta \eta = |\eta_1 -
\eta_W| - |\eta_2 - \eta_W|$.  Where $\eta_{1,2}$ are the rapidities
of the leptons ordered in the descending order of $p_T$.  We use the
difference from $\eta_W$ to make the variable invariant under Lorentz
boosts in the beam direction.  This variable is most likely to be
modified after taking into account initial and final state radiation
(ISR and FSR) effects as the number of jets are modified.  We shall
deal with this concern in Section \ref{sec:ISR}.

We also construct a similar variable, $\Delta |\eta|=|\eta_1| -
|\eta_2|$ which shows sensitivity to ${\rm Im}(\tilde b)$ and is much
less sensitive to ${\rm Re}(\tilde b)$.  It also has the added
advantage that one need not reconstruct the W and therefore can look
into inclusive SSD final states and is therefore expected to be more
robust to FSR effects.  However, it should be noted that this variable
is not invariant under longitudinal boosts.

\begin{figure}[htb]
\includegraphics[width=80mm]{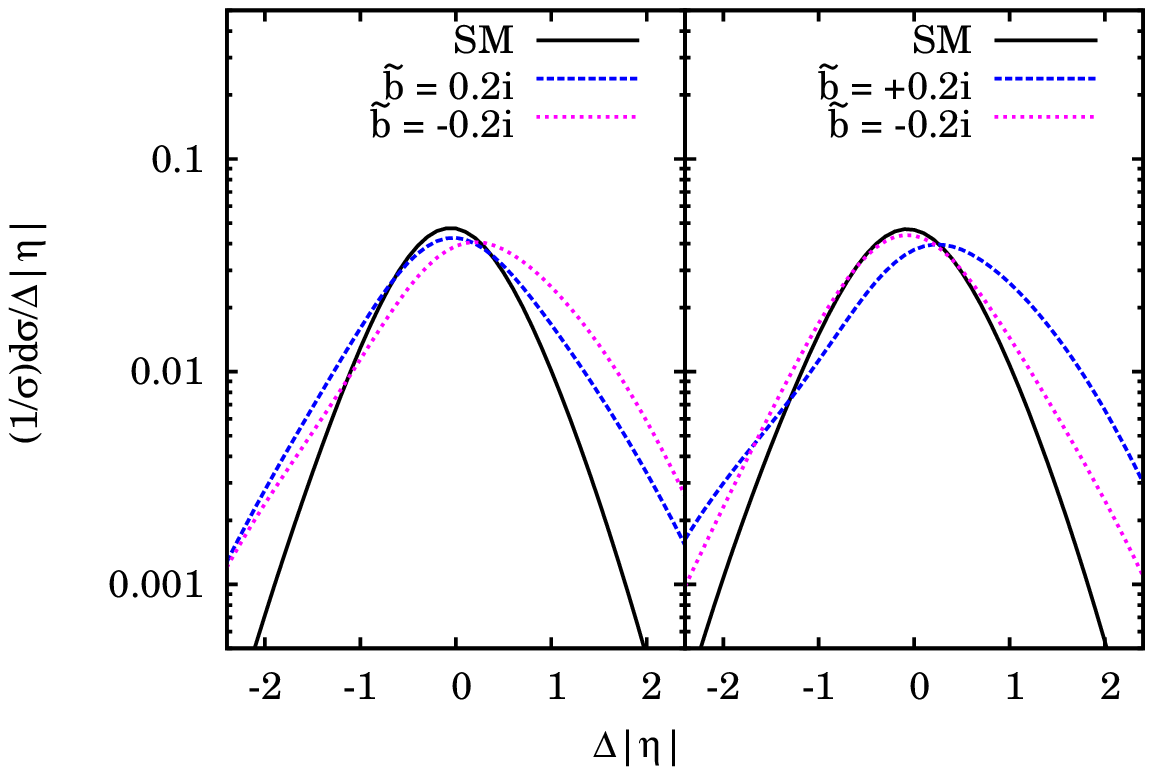}
\caption{\label{fig:detac-ssd} Effect of positive (dashed) and negative
  (dot-dashed) values of ${\rm Im}(\tilde b)$ on the $\Delta \eta$
  distributions for $m_H=150$~GeV.  The left panel corresponds to
  $\ell^+\ell^+$ and the right to $\ell^-\ell^-$ final states.}
\end{figure}

\begin{figure}[htb]
\includegraphics[width=80mm]{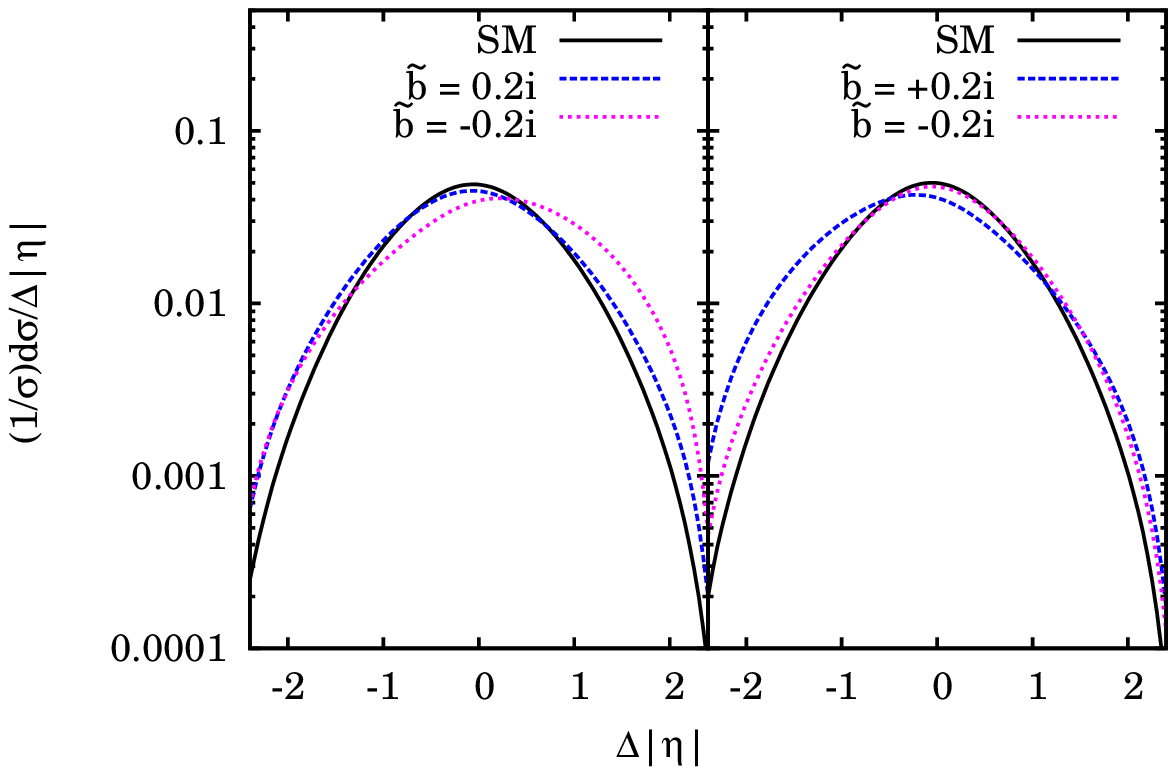}
\caption{\label{fig:dabsetac-ssd} Effect of positive (dashed) and
  negative (dot-dashed) values of ${\rm Im}(\tilde b)$ on the $\Delta
  |\eta|$ distributions for $m_H=150$~GeV.  The left panel corresponds
  to $\ell^+\ell^+$ and the right to $\ell^-\ell^-$ final states.}
\end{figure}

The distributions of $\Delta \eta$ and $\Delta |\eta|$ are shown in
Figure~\ref{fig:detac-ssd} and \ref{fig:dabsetac-ssd} respectively.
In both the cases, the $\ell^+\ell^+$ final state is particularly
sensitive to ${\rm Im}(\tilde b)<0$ whereas the $\ell^-\ell^-$ one is
sensitive to ${\rm Im}(\tilde b)>0$.  Therefore, we can use these
variables to confirm the presence of a non-zero ${\rm Im}(\tilde b)$
as only one of $\ell^+\ell^+$ or $\ell^-\ell^-$ will show a
significant deviation from the SM value.  The first variable is useful
because it shows a larger asymmetry and can therefore be used with
lower luminosity.  However, the shift in the curve is independent of
the sign of ${\rm Im}(\tilde b)$.  The second variable on the other
hand, has a lower asymmetry but changes sign depending on the sign of
${\rm Im}(\tilde b)$.  We also find that the effect of non-zero ${\rm
  Re}(\tilde b)$ is much smaller and is un-correlated with the its
sign.  Here too, we can construct left-right asymmetries to better
parametrise this difference.

\begin{eqnarray}
A_{SSD3}& = &\frac{\sigma(\Delta \eta>0) - \sigma(\Delta
  \eta<0)}{\sigma(\Delta \eta>0) + \sigma(\Delta \eta<0)}\\
A_{SSD4}& = &\frac{\sigma(\Delta |\eta|>0) - \sigma(\Delta
  |\eta|<0)}{\sigma(\Delta |\eta|>0) + \sigma(\Delta |\eta|<0)}
\end{eqnarray}

\begin{figure}[htb]
\includegraphics[width=80mm]{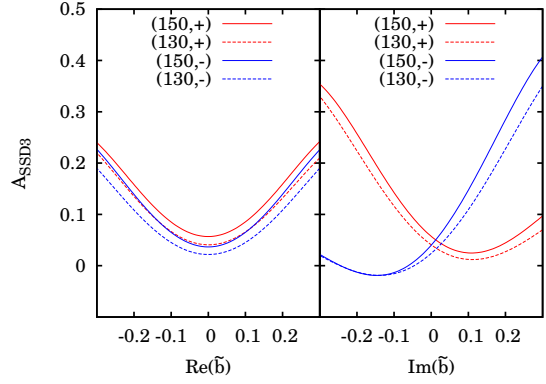}
\caption{\label{fig:asym-deta-lr} The asymmetry $A_{SSD3}$ for
  different values of the anomalous couplings.  The labels indicate
  the Higgs mass and the sign of the SSD.}
\end{figure}

\begin{figure}[htb]
\includegraphics[width=80mm]{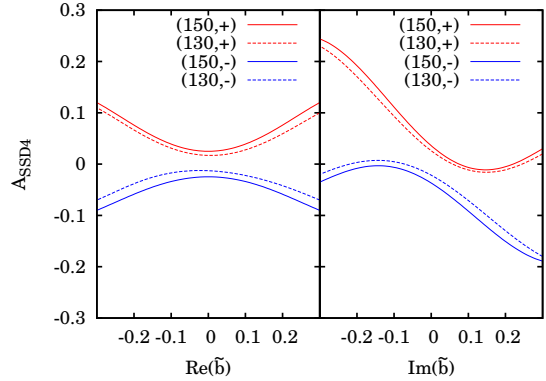}
\caption{\label{fig:asym-abs-deta-lr} The asymmetry $A_{SSD4}$ for
  different values of the anomalous couplings.  The labels indicate
  the Higgs mass and the sign of the SSD.}
\end{figure}

The distribution of the asymmetry $A_{SSD3}$ for different values of
${\rm Re}(\tilde b)$ and ${\rm Im}(\tilde b)$ is shown in
Figure~\ref{fig:asym-deta-lr}.  We can see that ${\rm Re}(\tilde b)$
affects both $\ell^+\ell^+$ or $\ell^-\ell^-$ symmetrically whereas
${\rm Im}(\tilde b)$ shows a very pronounced asymmetry depending on
sign.  For $\ell^+ \ell^+$ events observed with an integrated
luminosity of 30(50) fb$^{-1}$ and $\tilde b=-0.2i$, we get an
asymmetry $A_{SSD3}=0.288 \pm 0.061(0.241 \pm 0.070)$ for
$m_H=150(130)$ GeV with as compared to the SM value of $-0.01$ (same
for both Higgs masses).  The distribution of $A_{SSD4}$ is shown in
Figure~\ref{fig:asym-abs-deta-lr}.  The left panel shows the
dependence on ${\rm Re}(\tilde b)$.  The asymmetry distribution is
symmetric with respect to its sign but is of opposite sign for
$\ell^+\ell^+$ and $\ell^-\ell^-$ states.  The right panel shows the
effect of ${\rm Im}(\tilde b)$.  We see that in this case as well, the
sign of ${\rm Im}(\tilde b)$ causes pronounced asymmetry in either
$\ell^+\ell^+$ or $\ell^-\ell^-$ states.  This asymmetry can therefore
supplement the conclusions from $A_{SSD3}$.  For $\ell^+\ell^+$ states
with $\tilde b=-0.2i$ and 30(50) fb$^{-1}$ integrated luminosity,
$A_{SSD4}$ takes the values $0.217 \pm 0.062 (0.191 \pm 0.071)$ for
$m_H=150(130)$ GeV.

In all we find that the presence of anomalous couplings makes the
$\Delta p_T$ distribution harder and enhances the back-to-back region
in the $\cos \theta$ distribution.  Since the reliable construction of
the asymmetries requires accumulation of a large data set, we first
test the presence of anomalous couplings using these two
distributions.  We can then use the three asymmetry variables to for
positive and negative SSD to determine what kind of anomalous coupling
is present.  Let the labels $(+)$ and $(-)$ refer to the charge of the
SSD.  Then we can conclude the following:
\begin{itemize}
\item $A_{SSD1}(\pm)=0 \Rightarrow {\rm Re}(\tilde b)=0$
\item $A_{SSD1}(+) \neq 0 \Rightarrow {\rm Re}(\tilde b) \neq 0$;
  $sign({\rm Re}(\tilde b))=-sign(A_{SSD1}(+))$
\item $|A_{SSD3,SSD4}(+)|<|A_{SSD3,SSD4}(-)| \Rightarrow {\rm Im}(\tilde b)>0$
\item $|A_{SSD3,SSD4}(+)|>|A_{SSD3,SSD4}(-)| \Rightarrow {\rm Im}(\tilde b)<0$
\end{itemize}

Since the asymmetry variables listed above are not explicitly
CP-violating, it is possible that they might also be affected by the
presence of CP-conserving anomalous coupling $b$.  We therefore wish
to determine if it is possible to get similar results from a non-zero
value of $b$ and whether it is possible to distinguish the effect of
the two kinds of couplings.

We perform a similar calculation of $pp \rightarrow h \ell \nu$ and $h
\rightarrow \ell \nu j j $ using the $HWW$ vertex given in
equation~\ref{eqn:vertex} with $\tilde b=0$ instead.  The cross
section of Higgs production after including $b$ is then required to
also be within the Tevatron bounds.  This corresponds to a value of
$|b|\leq 0.05$ which will be used for the rest of this section.  We
then examine the three asymmetries defined in the previous section
with the same cuts.

\begin{figure}[htb]
\includegraphics[width=80mm]{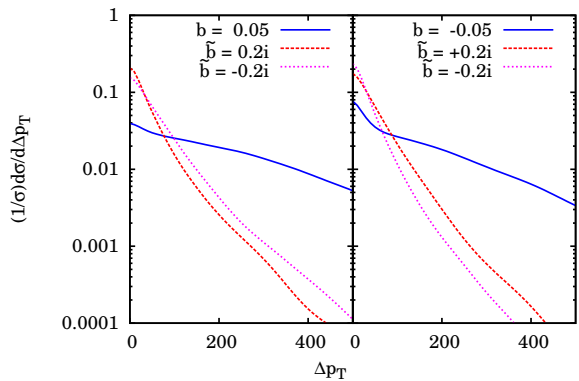}
\caption{\label{fig:pt-cpc} Comparison of the normalised $\Delta p_T$
  distribution between two same-sign leptons for $m_H=150$~GeV for
  values of $b$ and $\tilde b$.  The left(right) panel corresponds to
  $\ell^+\ell^+$ ($\ell^-\ell^-$)-type process.}
\end{figure}

We find that the $\Delta \phi$ asymmetry $A_{SSD1}$ and the $\Delta
|\eta|$-based $A_{SSD4}$ are both completely unaffected by the
presence of $b$.  Therefore, these two together can constitute robust
variables at the LHC for confirming the presence of a CP-violating
anomalous HWW coupling.  The second $\Delta \phi$-based asymmetry,
$A_{SSD2}$ is more negative in the case of CP-conserving anomalous
couplings.  However, the difference is small and measuring it with
accuracy will require a large luminosity.  The $\Delta \eta$-based
$A_{SSD3}$ shows similar behaviour between non zero values $b$ and
${\rm Im}(\tilde b)$.  We can further discriminate between $b$ or
$\tilde b$ type coupling by examining the $\Delta p_T$ distribution
which falls off much slower in the case of the CP conserving coupling.
This can set apart the presence of ${\rm Im}(\tilde b)$ quite
distinctly.  As an illustration, we present a comparison in
Figure~\ref{fig:pt-cpc}.  We find that difference in the distributions
for $b>0 (b<0)$ is probed best in $\ell^+\ell^+$ ($\ell^-\ell^-$)
channels irrespective of the sign of ${\rm Im}(\tilde b)$.  In both
cases, we find the distributions are distinct enough to allow us to
separate the effects from the two couplings.

\section{Effect of showering and hadronisation}
\label{sec:ISR}
Until now we have been working under the simplified scheme of
parton-level Monte-Carlo analysis.  However, initial and final-state
radiation play a very important role at the LHC.  In particular the
entire partonic system can acquire a transverse momentum due to recoil
from ISR.  One therefore needs to examine whether the effects of
showering destroy the correlations we had examined in the previous
section.  In this section, we investigate this in the context of the
distributions and asymmetries defined above.

We have started by obtaining unweighted events from the parton-level
code, which are then passed through
PYTHIA8\cite{Sjostrand:2007gs,Sjostrand:2006za} using the LHEF file
format\cite{Alwall:2006yp}.  PYTHIA8 performs the initial and final
state showers and hadronisation after which we use FastJet~2.4.1 with
the anti-kt algorithm\cite{Cacciari:2008gp} with a cone size parameter
of 0.4 to form the jets.  Leptons are considered isolated if the sum
of $E_T$ of particles around the lepton within a cone of 0.2 is less
than 10 GeV and the separation with the nearest jet is greater than
0.4.  All the variables and asymmetries are defined as before.

\begin{figure}[htb]
\includegraphics[width=80mm]{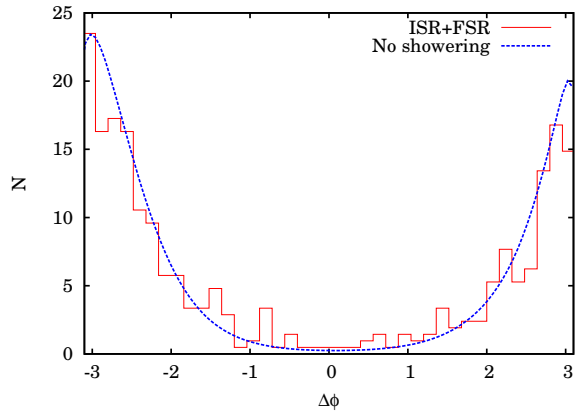}
\caption{\label{fig:compare-dphi}Comparison of the $\Delta \phi$
  distribution before and after including ISR and FSR for
  $\ell^+\ell^+$ final states and a value of $\tilde b=0.2$ and
  $m_H=150$ GeV for 30 fb$^{-1}$.}
\end{figure}

\begin{figure}[htb]
\includegraphics[width=80mm]{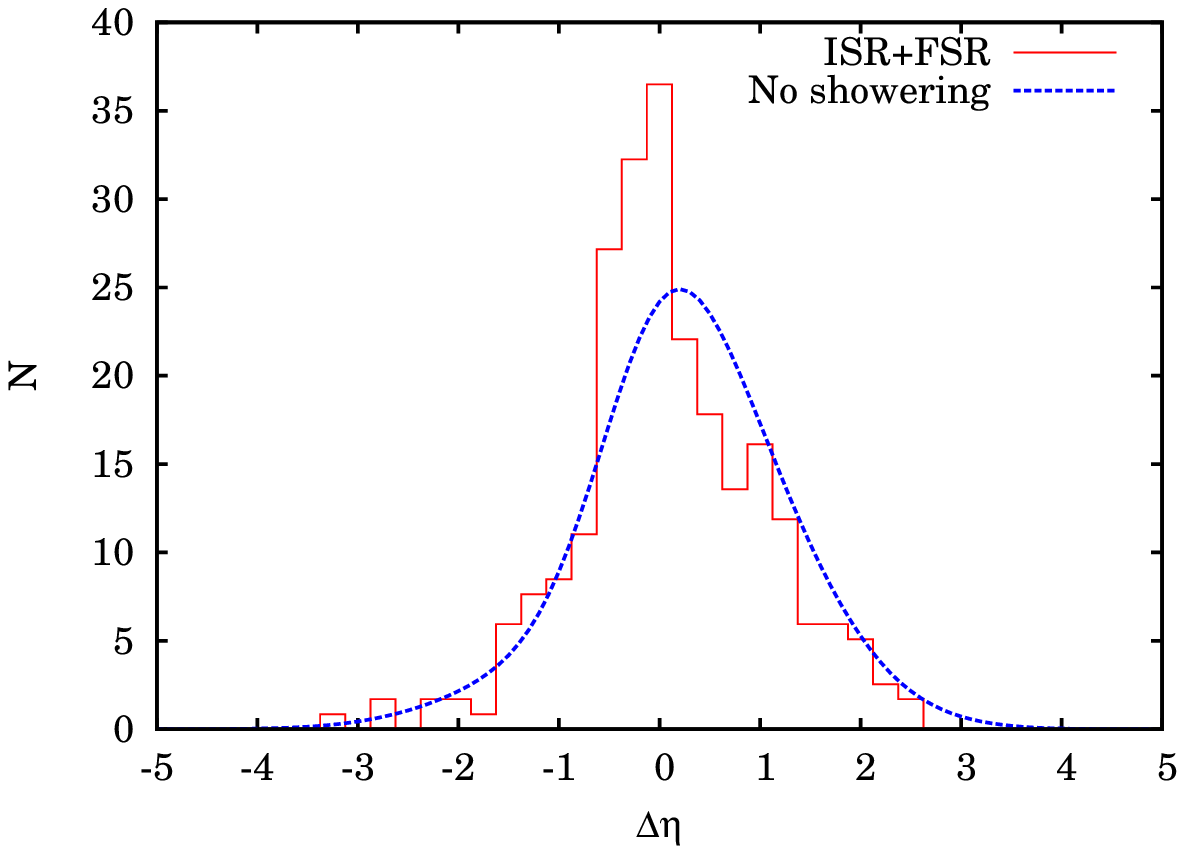}
\caption{\label{fig:compare-deta}Comparison of the $\Delta \eta$
  distribution before and after including ISR and FSR for
  $\ell^-\ell^-$ final states and a value of $\tilde b=0.2i$ and
  $m_H=150$ GeV for 30 fb$^{-1}$.}
\end{figure}

\begin{figure}[htb]
\includegraphics[width=80mm]{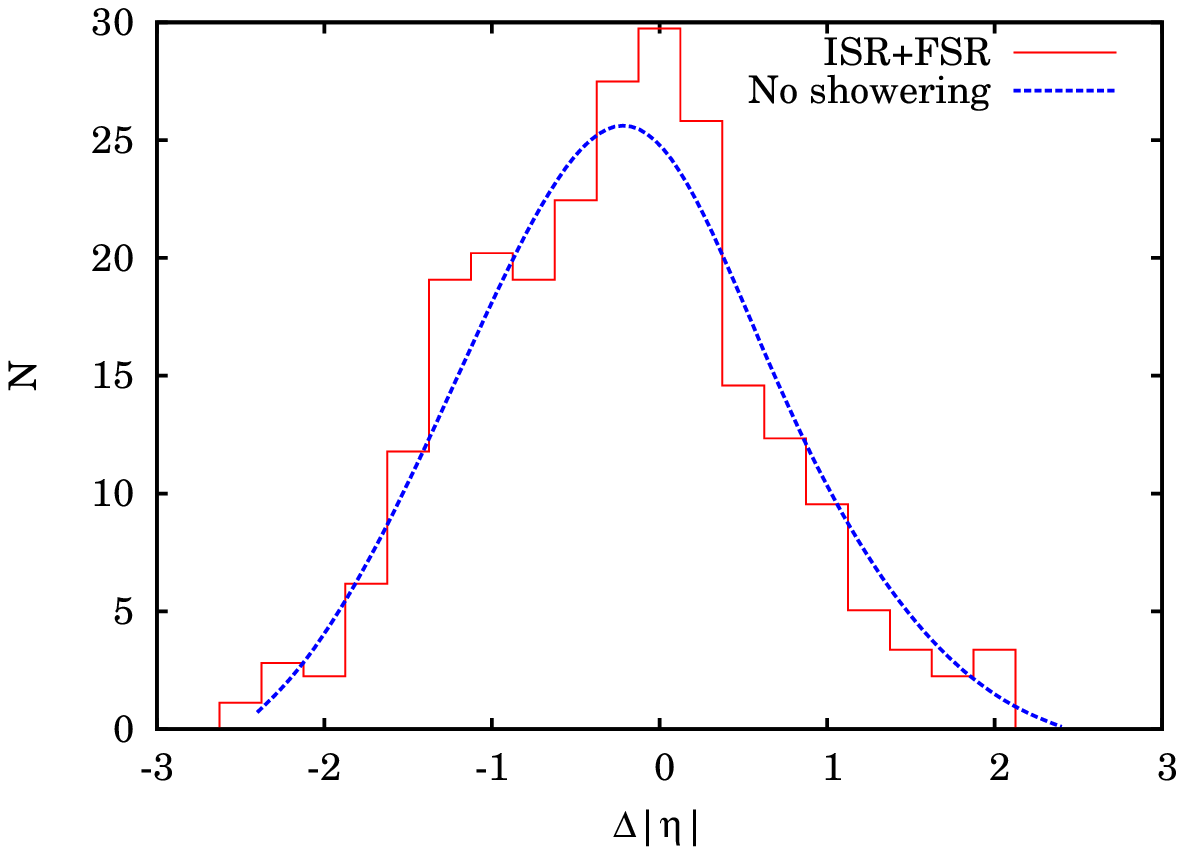}
\caption{\label{fig:compare-abs-deta}Comparison of the $\Delta |\eta|$
  distribution before and after including ISR and FSR for
  $\ell^-\ell^-$ final states and a value of $\tilde b=0.2i$ and
  $m_H=150$ GeV for 30 fb$^{-1}$.}
\end{figure}

As an illustration, we first present the $\Delta \phi$ distributions
for a $\ell^+\ell^+$ final state for a value of $\tilde b=0.2$ in
Figure \ref{fig:compare-dphi}.  It can be seen that the distribution
retains the correct left-right asymmetry.  The $\Delta \eta$
distribution for $\ell^-\ell^-$ and a value of $\tilde b=0.2i$ is
shown in Figure \ref{fig:compare-deta} and the $\Delta |\eta|$
distribution is shown in Figure \ref{fig:compare-abs-deta}.  In these
cases too, we see that the distribution is fairly unchanged.  Both
these distributions can therefore be thought of as a robust variables
for LHC analyses.

We also present the values of the asymmetry variable constructed in
the previous sections in Table \ref{tab:asym-ISR}.  The variable
$A_{SSD1}$ is the most robust as the values change only very slightly.
The $\Delta \eta$ dependent $A_{SSD3}$ still shows an asymmetry based
on sign of ${\rm Im}(\tilde b)$ but the effect is diluted after taking ISR
effects into account.

\begin{table}
\begin{tabular}{|ll|r|r|r|r|r|}
\hline
          &         &  $\tilde b=0$   &   $\tilde b=0.2$  & $\tilde b=-0.2$  &  $\tilde b=0.2i$  &  $\tilde b=-0.2i$  \\
\hline
$A_{SSD1}$  &  $W^+$  &   0.03(0.00)  &  -0.27(-0.21)  &   0.19(0.21)  &  0.08(0.00)  & -0.07(0.00)  \\
           &  $W^-$  &  -0.08(0.00)  &   0.27(0.20)   &  -0.19(-0.20) & -0.03(0.00)  &  0.03(0.00)  \\
\hline
 $A_{SSD2}$  &  $W^+$  &  -0.73(-0.79)  &  -0.82(-0.89)  &  -0.80(-0.87)  &  -0.76(-0.88)  &  -0.77(0.87)  \\
            &  $W^-$  &  -0.61(-0.77)  &  -0.78(-0.86)  &  -0.80(-0.86)  &  -0.78(-0.88)  &  -0.70(-0.83)  \\
\hline
$A_{SSD3}$  &  $W^+$  &   0.05(-0.01)  &   0.06(0.17)  &   0.06(0.17)  &   0.03(0.04)  &   0.12(0.31)  \\
           &  $W^-$  &   0.02(-0.03)  &   0.10(0.15)  &   0.06(0.15)  &   0.12(0.29)  &   0.03(0.02)  \\
\hline
$A_{SSD4}$  &  $W^+$  &   -0.01(-0.01)  &    0.02(0.08)  &   0.04(0.08)  &  -0.01(-0.01)  &   0.13(0.22)  \\
           &  $W^-$  &   -0.05(-0.01)  &  -0.08(-0.06)  & -0.14(-0.06)  &  -0.11(-0.17)  &  -0.03(-0.01)  \\

\hline
\end{tabular}
\caption{\label{tab:asym-ISR} Asymmetries after ISR and FSR for $m_H=$
  150 GeV.  The value from parton-level calculations is given in the
  parentheses for comparison.}
\end{table}

\section{Conclusions}
\label{sec:conclusions}
We have systematically examined the effects of a CP-violating HWW
coupling on Higgs production and decay at the LHC.  We probe this
coupling via the $WH$ associated production followed by $H\rightarrow
WW^*\rightarrow \ell \nu f \bar f'$ which gives rise to same-sign
dilepton final states.  We take into account the Tevatron limits on
the Higgs cross section to restrict the values of real and imaginary
parts of the anomalous coupling.  We find that, besides enhancing the
production cross section, it also causes significant deviations in
various kinematic correlations between leptons in the final state.

We have presented several variables whose distributions show
significant deviation from the standard model case.  We also define
asymmetries constructed from three of them, viz. $\Delta \phi$,
$\Delta \eta$ and $\Delta |\eta|$, which can show significant
deviation from SM predictions.  Trends in the $\Delta p_T$ and $\cos
\theta$ distributions may be used to first ascertain the presence of
an anomalous coupling.  The left-right asymmetry in the $\Delta \phi$,
$\Delta \eta$ and $\Delta |\eta|$ distributions can be used to probe
its nature in detail.  After imposing cuts required to suppress the SM
backgrounds, the asymmetries can be discerned at the 3(5)$\sigma$
level at 14 TeV, with an integrated luminosity of 30(50) fb$^{-1}$ for
a Higgs of mass 150 TeV.  The asymmetries for a Higgs mass of 130 GeV
can be similarly determined at 3(5) sigma with 50(140) fb$^{-1}$.  Its
should be noted that our calculation is done at the leading order, and
the inclusion of an appropriate next-to-leading order K-factor is
expected to enhance the signal rates.  We also present and compare
various distributions at the parton level and after showering and
hadronisation.  We find that our conclusions are largely unchanged,
even after taking the latter effects into account.

\begin{acknowledgments}
 ND and BM are partially supported by funding available from the
 Department of Atomic Energy, Government of India for the Regional
 Centre for Accelerator-based Particle Physics(RECAPP), Harish-Chandra
 Research Institute.  DKG acknowledges partial support from the
 Department of Science and Technology, India under the grant
 SR/S2/HEP-12/2006.  ND and BM also thank the Indian Association for
 the Cultivation of Science for hospitality at the initial stage of
 this work.  DKG acknowledges the hospitality the Abdus Salam
 International Centre for Theoretical Physics and RECAPP at a later
 stage of the project.  Computational work for this study was
 partially carried out at the cluster computing facility of
 Harish-Chandra Research Institute
 (\url{http://cluster.mri.ernet.in}).
\end{acknowledgments}

\bibliography{paper}
\end{document}